\documentclass{sigchi}

\CopyrightYear{2017} 
\setcopyright{acmlicensed}
\conferenceinfo{CSCW '17,}{February 25-March 01, 2017, Portland, OR, USA}
\isbn{978-1-4503-4335-0/17/03}\acmPrice{\$15.00}
\doi{http://dx.doi.org/10.1145/2998181.2998248}

\usepackage{balance}  
\usepackage{graphics} 
\usepackage[T1]{fontenc}
\usepackage{txfonts}
\usepackage{times}    
\usepackage[pdftex]{hyperref}
\usepackage{url}      
\usepackage{color}
\usepackage{textcomp}
\usepackage{booktabs}
\usepackage{ccicons}
\usepackage{subcaption}

\usepackage{color}
\usepackage[dvipsnames]{xcolor}
\usepackage{cite}
\usepackage{multirow}
\usepackage{siunitx}

\newcommand{\first}[1]{\subsection{#1}}
\newcommand{\second}[1]{\textit{#1}}

\makeatletter
\def\url@leostyle{%
  \@ifundefined{selectfont}{\def\UrlFont{\sf}}{\def\UrlFont{\small\bf\ttfamily}}}
\makeatother
\def\pprw{8.5in}
\def\pprh{11in}

\definecolor{linkColor}{RGB}{6,125,233}
\hypersetup{%
  pdftitle={SIGCHI Conference Proceedings Format},
  pdfauthor={Kenji Hata, Ranjay Krishna, Li Fei-Fei, Michael S. Bernstein},
  pdfkeywords={Crowdsourcing; microtask; fatigue; consistent; satisficing},
  bookmarksnumbered,
  pdfstartview={FitH},
  colorlinks,
  citecolor=black,
  filecolor=black,
  linkcolor=black,
  urlcolor=linkColor,
  breaklinks=true,
}

\begin{document}

\title{A Glimpse Far into the Future: Understanding Long-term Crowd Worker Quality}

\numberofauthors{1}
\author{\alignauthor{Kenji Hata, Ranjay Krishna, Li Fei-Fei, Michael S. Bernstein}\\
\affaddr{Stanford University}\\
\email{\{kenjihata, ranjaykrishna, feifeili, msb\}@cs.stanford.edu}}

\maketitle

\begin{abstract}
 Microtask crowdsourcing is increasingly critical to the creation of extremely large datasets. As a result, crowd workers spend weeks or months repeating the exact same tasks, making it necessary to understand their behavior over these long periods of time. 
 We utilize three large, longitudinal datasets of nine million annotations collected from Amazon Mechanical Turk to examine claims that workers fatigue or satisfice over these long periods, producing lower quality work.
 We find that, contrary to these claims, workers are extremely stable in their quality over the entire period.
 To understand whether workers set their quality based on the task's requirements for acceptance, we then perform an experiment where we vary the required quality for a large crowdsourcing task.
 Workers did not adjust their quality based on the acceptance threshold: workers who were above the threshold continued working at their usual quality level, and workers below the threshold self-selected themselves out of the task.
 Capitalizing on this consistency, we demonstrate that it is possible to predict workers' long-term quality using just a glimpse of their quality on the first five tasks.
\end{abstract}



\section{Introduction}
Microtask crowdsourcing is gaining popularity among corporate and research communities as a means to leverage parallel human computation for extremely large problems~\cite{thomee2015new,bernstein2010soylent,bernstein2011crowds,lasecki2011real,lasecki2012real}. These communities use crowd work to complete hundreds of thousands of tasks per day~\cite{marcuswaran}, from which new datasets with over $20$ million annotations can be produced within a few months~\cite{krishnavisualgenome}. 
A crowdsourcing platform like Amazon's Mechanical Turk (AMT) is a marketplace subject to human factors that affect its performance, both in terms of speed and quality~\cite{difallah2015dynamics}. 
Prior studies found that work division in crowdsourcing follows a Pareto principle, where a small minority of workers usually completes a great majority of the work~\cite{little8020}. If such large crowdsourced projects are being completed by a small percentage of workers, then these workers spend hours, days, or weeks executing the exact same tasks. Consequently, we pose the question:

\textit{How does a worker's quality change over time?}

Multiple arguments from previous literature in psychology suggest that quality should decrease over time. \emph{Fatigue}, a temporary decline in cognitive or physical condition, can gradually result in performance drops over long periods of time~\cite{perelli1980fatigue, boksem2008mental,krueger1989sustained}. Since the microtask paradigm in large scale crowdsourcing involves monotonous sequences of repetitive tasks, fatigue buildup can pose a potential problem to the quality of submitted work over time~\cite{dai2015and}. Furthermore, workers have been noted to be ``satisficers'' who, as they gain familiarity with the task and its acceptance thresholds, strive to do the minimal work possible to achieve these thresholds~\cite{simon1972theories, chandler2013risks}.

To study these long term effects on crowd work, we analyze worker trends over three different real-world, large-scale datasets~\cite{krishnavisualgenome} collected from microtasks on AMT: image descriptions, question answering, and binary verifications. With microtasks comprising over $60\%$ of the total crowd work and microtasks involving images being the most common type~\cite{pew2016}, these datasets cover a large percentage of the type of crowd work most commonly seen. Specifically, we use over $5$ million image descriptions from \num[group-separator={,}]{2674} workers over a $9$ month span, $0.8$ million question-answer pairs from \num[group-separator={,}]{2179} workers over a $3$ month span, and $2$ million verifications from \num[group-separator={,}]{3913} workers over a $3$ month span. The average worker in the largest dataset worked for an average of $2$ eight-hour work days while the top $1\%$ of workers worked for nearly $45$ eight-hour work days. Using these datasets, we look at temporal trends in the accuracy of annotations from workers, diversity of these annotations, and the speed of completion.

Contrary to our hypothesis that workers would exhibit glaring signs of fatigue via large declines in submission quality over time, we find that workers who complete large sets of microtasks maintain a consistent level of quality (measured as the percentage of correct annotations). Furthermore, as workers become more experienced on a task, they develop stable strategies that do not change, enabling them to complete tasks faster. But are workers generally consistent or is this consistency simply a product of the task design?

We thus perform an experiment where we hire workers from AMT to complete large-scale tasks while randomly assigning them into different task designs. These designs were varied across two factors: the acceptance \textit{threshold} with which we accept or reject work, and the \textit{transparency} of that threshold. If workers manipulate their quality level  strategically to avoid rejection, workers with a high (difficult) threshold would perform at a noticeably better level than the ones with a low threshold who can satisfice more aggressively. However, this effect might only be easily visible if workers have transparency into how they performed on the task.

By analyzing \num[group-separator={,}]{676628} annotations collected from \num[group-separator={,}]{1134} workers in the experiment on AMT, we found that workers display consistent quality regardless of their assigned condition, and that lower-quality workers in the high threshold condition would often self-select out of tasks where they believe there is a high risk of rejection. Bolstered by this consistency, we ask: can we predict a worker's future quality months after they start working on a microtask?

If individual workers indeed sustain constant correctness over time, then, intuitively, any subset of a worker's submissions  should be representative of their entire work.  We demonstrate that a simple glimpse of a worker's quality in their first few tasks is a strong predictor of their long-term quality. Simply averaging the quality of work of a worker's first $5$ completed tasks can predict that worker's quality during the final $10\%$ of their completed tasks with an average error of $3.4\%$.

Long-term worker consistency suggests that paying attention to easy signals of good workers can be key to collecting a large dataset of high quality annotations~\cite{mitra2015comparing,rzeszotarski2012crowdscape}. Once we have identified these workers, we can back off the gold-standard (attention check) questions to ensure good quality work, since work quality is unvarying~\cite{liueffective}. We can also be more permissive about errors from workers known to be good, reducing the rejection risk that workers face and increasing worker retention~\cite{difallah2014scaling,law2016curiosity}.

\section{Related Work}
Our work is inspired by psychology, decision making, and workplace management literature that focuses on identifying the major factors that affect the quality of work produced. Specifically, we look at the effects of fatigue and satisficing in the workplace. We then study whether these problems transfer to the crowdsourcing domain. Next, we explore how our contributions are necessary to better understand the global ecosystem of crowdsourcing. Finally, we discuss the efficacy of existing worker quality improvement techniques.

\first{Fatigue}
Repeatedly completing the same task over a sustained period of time will induce fatigue, which increases reaction time, decreases production rate, and is linked to a rise in poor decision-making~\cite{krueger1989sustained, wyatt1937fatigue}.  The United States Air Force found that both the cognitive performance and physical conditions of its airmen continually deteriorated during the course of long, mandatory shifts~\cite{perelli1980fatigue}. However, unlike these mandatory, sustained shifts,  crowdsourcing is generally opt-in for workers --- there always exists the option for workers to break or find another task whenever they feel tired or bored~\cite{lasecki2014using,lasecki2015effects}. Nonetheless, previous work has shown that people cannot accurately gauge how long they need to rest after working continuously, resulting in incomplete recoveries and drops in task performance after breaks~\cite{hennfng1989microbreak}. Ultimately, previous work in fatigue suggests that crowd workers who continuously complete tasks over sustained periods would result in significant decreases in work quality. We show that contrary to this literature, crowd workers remain consistent throughout their time on a specific task.

\first{Satisficing}
Crowd workers are often regarded as ``satisficers'' who do the minimal work needed for their work to be accepted~\cite{simon1972theories,chandler2013risks}. Examples of satisficing in crowdsourcing occur during surveys \cite{krosnick1991response} and when workers avoid the most difficult parts of a task~\cite{mason2010financial}. Disguised attention checks in the instructions~\cite{oppenheimer2009instructional} or rate-limiting the presentation of the questions~\cite{kapelner2010preventing} improves the detection and prevention of satisficing. Previous studies of crowd workers' perspectives find that crowd workers believe themselves to be genuine workers, monitoring their own work and giving helpful feedback to requesters~\cite{mcinnis2016taking}. Workers have also been shown to respond well and produce high quality work if the task is designed to be effort-responsive~\cite{ho2015incentivizing}. However, workers often consider the cost-benefit of continuing to work on a particular task --- if they feel that a task is too time-consuming relative to its reward, then they often drop out or compensate by satisficing (e.g. reducing quality)~\cite{mcinnis2016taking}. Prior work has shown that We observe that satisficing does occur, but it only affects a small portion of long-term workers. We also observe in our experiments that workers opt out of tasks where they feel they have a high risk of rejection.

\first{The global crowdsourcing ecosystem}
With the rapidly growing size of crowdsourcing projects, workers now have the opportunity to undertake large batches of tasks. As they progress through these tasks, questions arise and they often seek help by communicating with other workers or the task creator \cite{martin2014being}. Furthermore, on external forums and in collectives, workers often share well-paying work opportunities, teach and learn from other workers, review requesters, and even consult with task creators to give constructive feedback~\cite{martin2014being,irani2013turkopticon,salehi2015we,mcinnis2016taking}. When considering this crowdsourcing ecosystem, crowd researchers often envision how more complex workflows can be integrated to make the overall system more efficient, fair, and allow for a wider range of tasks to be possible~\cite{kittur2013future}. To continue the trend towards a more complex, but more powerful, crowdsourcing ecosystem, it is imperative that we study the long-term trends of how workers operate within it. Our paper seeks to identify trends that occur as workers continually complete tasks over a long period of time. We conclude that crowdsourcing workflows should design methods to identify good workers and provide them with the ability to complete tasks with a low threshold for acceptance as good workers work consistently hard regardless of the acceptance criteria.

\begin{figure*}[t!]
    \centering
    \includegraphics[width=\textwidth]{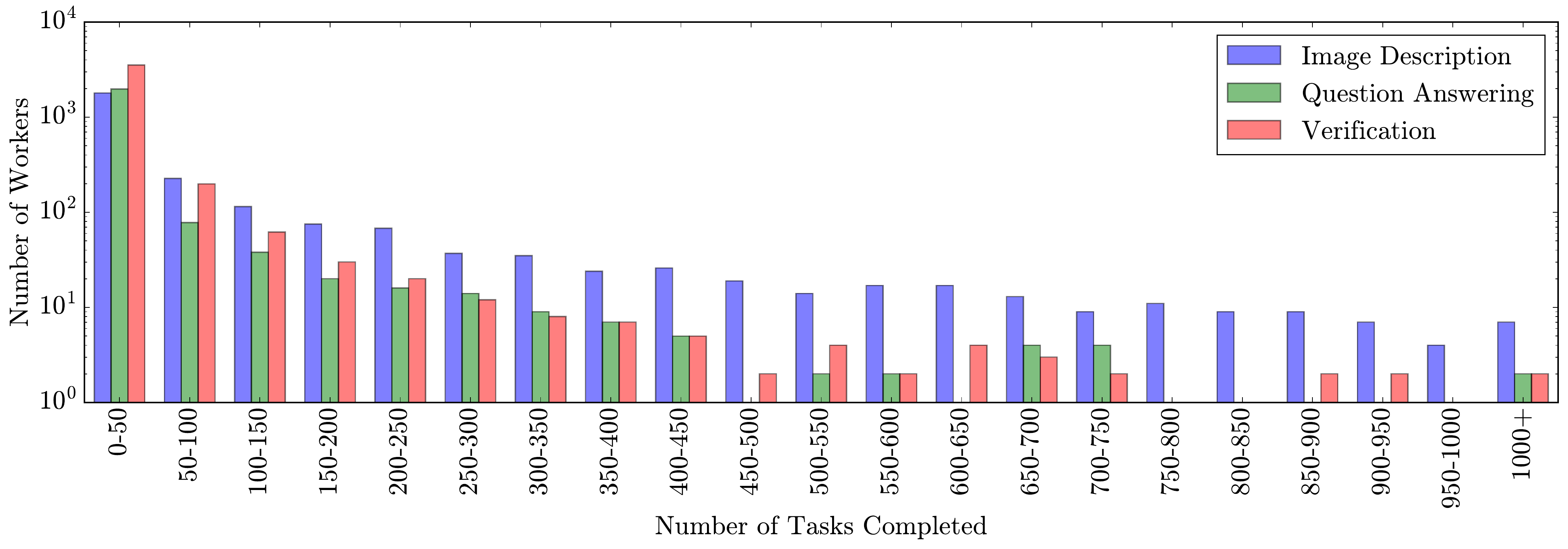}
    \caption{A distribution of the number of workers for each of the three datasets. A small number of persistent workers complete most of the work: the top $20\%$ of workers completed roughly $90\%$ of all tasks.}
    \label{fig:dist_workers}
\end{figure*}

\first{Improving crowdsourcing quality}
External checks such as verifiable gold standards, requiring explanations, and majority voting are standard practice for reducing bad answers and quality control~\cite{kittur2008crowdsourcing,callison2009fast}. Other methods directly estimate worker quality to improve these external checks~\cite{ipeirotis2010quality,whitehill2009whose}. Giving external feedback or having crowd workers internally reflect on their prior work also has been shown to yield better results~\cite{dow2012shepherding}. Previous work directly targets the monotony of crowdsourcing, showing that by framing the task as more meaningful to workers (for example as a charitable cause), one obtains higher quality results~\cite{chandler2013breaking}. However, this framing study only had workers do each task a few times and did not observe long-term trends. We, on the other hand, explore the changes in worker quality on microtasks that are repeated by workers over long periods of time.

\section{Analysis: Long-Term Crowdsourcing Trends}
In this section, we perform an analysis of worker behavior over time on large-scale datasets of three machine learning labeling tasks: image descriptions, question answering, and binary verification. We examine common trends, such as worker accuracy and annotation diversity over time. We then use our results to answer whether workers are fatiguing or displaying other decreases in effectiveness over time. 

\first{Data}
We first describe the three datasets that we inspect. Each of the three tasks were priced such that workers could earn $\$6$ per hour and were only available to workers with a $95\%$ approval rating and who live in the United States. For the studies in this paper, workers were tracked by their AMT worker ID's. The tasks and interfaces used to collect the data are described in further detail in the Visual Genome paper~\cite{krishnavisualgenome}. 

\second{Image descriptions.}
An image description is a phrase or sentence associated with a certain part of an image. To complete this task, a worker looks at an image, clicks and drags to select an area of the image, and then describes it using a short textual phrase (e.g., ``The dog is jumping to catch the frisbee''). Each image description task requires a worker to create $5-10$ unique descriptions for one randomly selected image, averaging at least $5$ words per description. Workers were asked to keep the descriptions factual and avoid submitting any speculative phrases or sentences. We estimate that each task takes around $4$ minutes and we allotted $2.5$ hours such that workers did not feel pressured for time. In total, $\num[group-separator={,}]{5380263}$ image descriptions were collected from $\num[group-separator={,}]{2674}$ workers over $9$ months.

\second{Question answers.}
Each question answering task asks a worker to write $7$ questions and their corresponding answers per image for $2$ different, randomly selected images. Workers were instructed to begin each sentence with one of the following questions: who, what, when, where, why and how~\cite{kuhn2013political}. Furthermore, to ensure diversity of question types, workers were asked to write a minimum of $4$ of these question types. Workers were also instructed to be concise and unambiguous to avoid wordy and speculative questions. Each task takes around $4$ minutes and we allotted $2.5$ hours such that workers did not feel pressured for time. In total, $\num[group-separator={,}]{832880}$ question-answer pairs were generated by $\num[group-separator={,}]{2179}$ workers over $3$ months.

\second{Binary verifications.}
Verification tasks were quality control tasks: given an image and a question-answer pair, workers were asked if the question was relevant to the image and if the answer accurately responded to the question. The majority decision of 3 workers was used to determine the accuracy of each question answering pair. For each verification task, a worker voted on $50$ randomly-ordered question-answer pairs. Each task takes around $3$ minutes and we allotted $1$ hour such that workers did not feel pressured for time. In total, $\num[group-separator={,}]{2498640}$ votes were cast by $\num[group-separator={,}]{3913}$ workers over $3$ months. 

\second{Overall.} 
Figure~\ref{fig:dist_workers} shows the distribution of how many tasks workers completed over the span of the data collection period, while Table~\ref{tab:dataset} outlines the total number of annotations and tasks completed. The top $20\%$ of workers who completed the most tasks did $91.8\%$, $90.9\%$, and $88.9\%$ of the total work in each of the three datasets respectively. These distributions are similar to the standard Pareto $80$-$20$ rule~\cite{little8020}, clearly demonstrating that a small, but persistent minority of workers completes an extremely large number of similar tasks. We noticed that workers in the top $1\%$ each completed approximately $1\%$ of the respective datasets each, with $\num[group-separator={,}]{5455}$ image description tasks, $758$ question answering tasks, and $\num[group-separator={,}]{1018}$ verification tasks completed on average. If each of these workers in the top $1\%$ took $4$ minutes for image descriptions and question answering tasks and $3$ minutes for verification tasks, the estimated average work time equates to $45$, $6.2$ and $6.2$ eight-hour work days for each task respectively. This sheer workload demonstrates that workers may work for very extended periods of time on the same task. Additionally, workers, on average, completed at least one task per week for $6$ weeks. By the final week of the data collection, about $10\%$ of the workers remained working on the tasks, suggesting that our study captures the entire lifetime of many of these workers. 

\begin{table}[t]
    \centering
    \begin{tabular}{l | r r r}
        & Annotations & Tasks & Workers \\ \hline
         \rule{0pt}{2ex}Descriptions & $\num[group-separator={,}]{5380263}$& $\num[group-separator={,}]{605443}$ & $\num[group-separator={,}]{2674}$ \\ 
         Question Answering & $\num[group-separator={,}]{830625}$ & $\num[group-separator={,}]{54587}$ & $\num[group-separator={,}]{2179}$ \\
         Verification & $\num[group-separator={,}]{2689350}$ & $\num[group-separator={,}]{53787}$  & $\num[group-separator={,}]{3913}$ \\ 
         
    \end{tabular}
    \caption{The number of workers, tasks, and annotations collected for image descriptions, question answering, and verifications.}
    \label{tab:dataset}
\end{table} 
We focus our attention on workers who completed at least $100$ tasks during the span of the data collection. The completion time for $100$ tasks is approximately $6.7$ hours for image description and question answering tasks and $5.0$ hours for verification tasks. We find that $657$, $128$, and $177$ workers completed $100$ of the image description, question answering, and verification tasks respectively. The median worker in each task type completed $349$, $220$, and $181$ tasks, which translates to $23.2$, $14.6$, and $6.0$ hours of continuous work.  These workers also produced $94.5\%$, $70.5\%,$ and $66.3\%$ of each of the total annotations. These worker pools are relatively unique: there are $61$ shared workers between image descriptions and QA, $69$ shared workers between image description and verification, $42$ shared workers between question answering and verifications, and $25$ shared workers between all three tasks.

\begin{figure}[t]
    \centering
    \includegraphics[width=0.45\columnwidth]{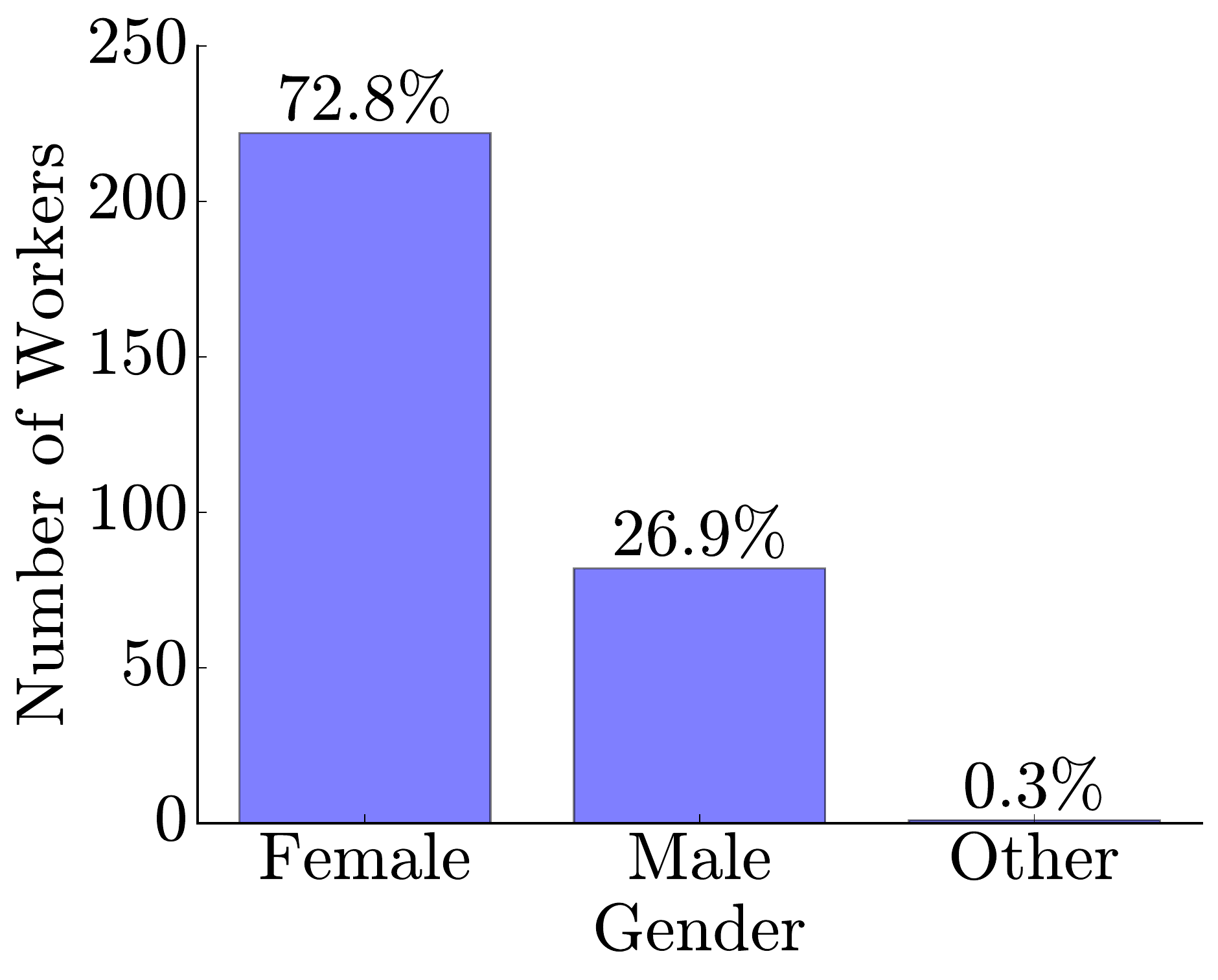}
    \includegraphics[width=0.45\columnwidth]{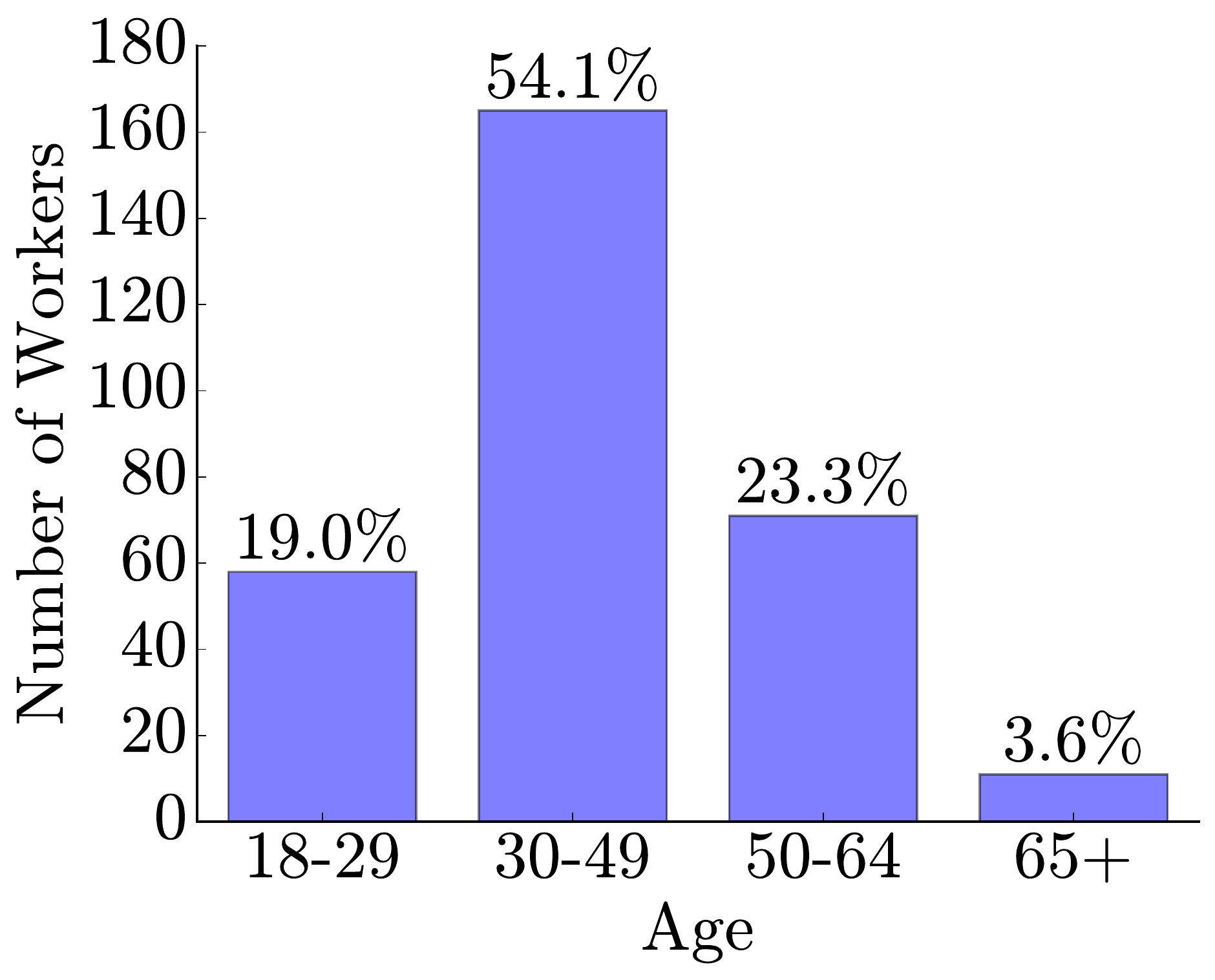}
    \caption{Self reported gender (left) and age distribution (right) of 298 workers who completed at least $100$ of the image description, question answer, or binary verification tasks.}
    \label{fig:survey}
\end{figure}

We reached out to the $815$ unique workers who had worked on at least $100$ tasks and asked them to complete a survey. After collecting $305$ responses, we found the gender distribution to be $72.8\%$ female, $26.9\%$ male, and $0.3\%$ other (Figure~\ref{fig:survey}). Furthermore, we found that workers with ages $30$-$49$ were the majority at $54.1\%$ of the long-term worker population. Ages $18$-$29$, $50$-$64$, and $65+$ respectively comprised $19.0\%$, $23.3\%$ and $3.6\%$ of the long-term worker population.  Compared to the distributions in previously gathered demographics on AMT~\cite{pew2016, difallah2014scaling,ross2010crowdworkers}, the gender and age distribution of all workers closely aligns with these other previously gathered distributions~\cite{krishnavisualgenome}. However, the distribution of long-term workers is skewed towards older and female workers. 

\first{Workers are consistent over long periods}
We analyzed worker accuracy and annotation diversity over the entire period of time that they worked on these tasks. Because workers performed different numbers of tasks, we normalize time data to percentages of their total \textit{lifetime}, which we define as the period from when a worker starts the task until they stop working on that task. For example, if one worker completed $200$ tasks and another completed $400$ tasks, then the halfway point in their respective lifetimes would be when they completed $100$ and $200$ tasks. 

\second{Annotation accuracy.} A straightforward metric of quality is the percentage of microtasks that are correct. To determine accuracy for an image description or question answering task, we computed the percentage of descriptions or question-answer pairs deemed true by a majority vote made by other workers. However, to use this majority vote in a metric, we need to first validate that this verification process is repeatable and accurate. Since the ground truth of verification tasks is unknown at such a large scale, we need a method to estimate the accuracy of each verification decision. We believe that comparing a worker's vote against the majority decision is a good approximation of accuracy. To test accuracy, we randomly sampled a set of $1,000$ descriptions and image answers and manually compared our own verifications against the majority vote, which resulted in a $98.2\%$ match. To test repeatability, we randomly sampled a set of $15,000$ descriptions and question answers to be sent back to be voted on by $3$ new workers $6$ months after the initial dataset was collected. Ultimately, we found a $99.3\%$ similarity between the majority decision of this new verification process with the original decision reported in the dataset~\cite{krishnavisualgenome}. The result of this test indicates that the majority decision is both accurate and repeatable, making it a good standard to compare against. 

\begin{figure}[t]
    \centering
    \includegraphics[width=\columnwidth]{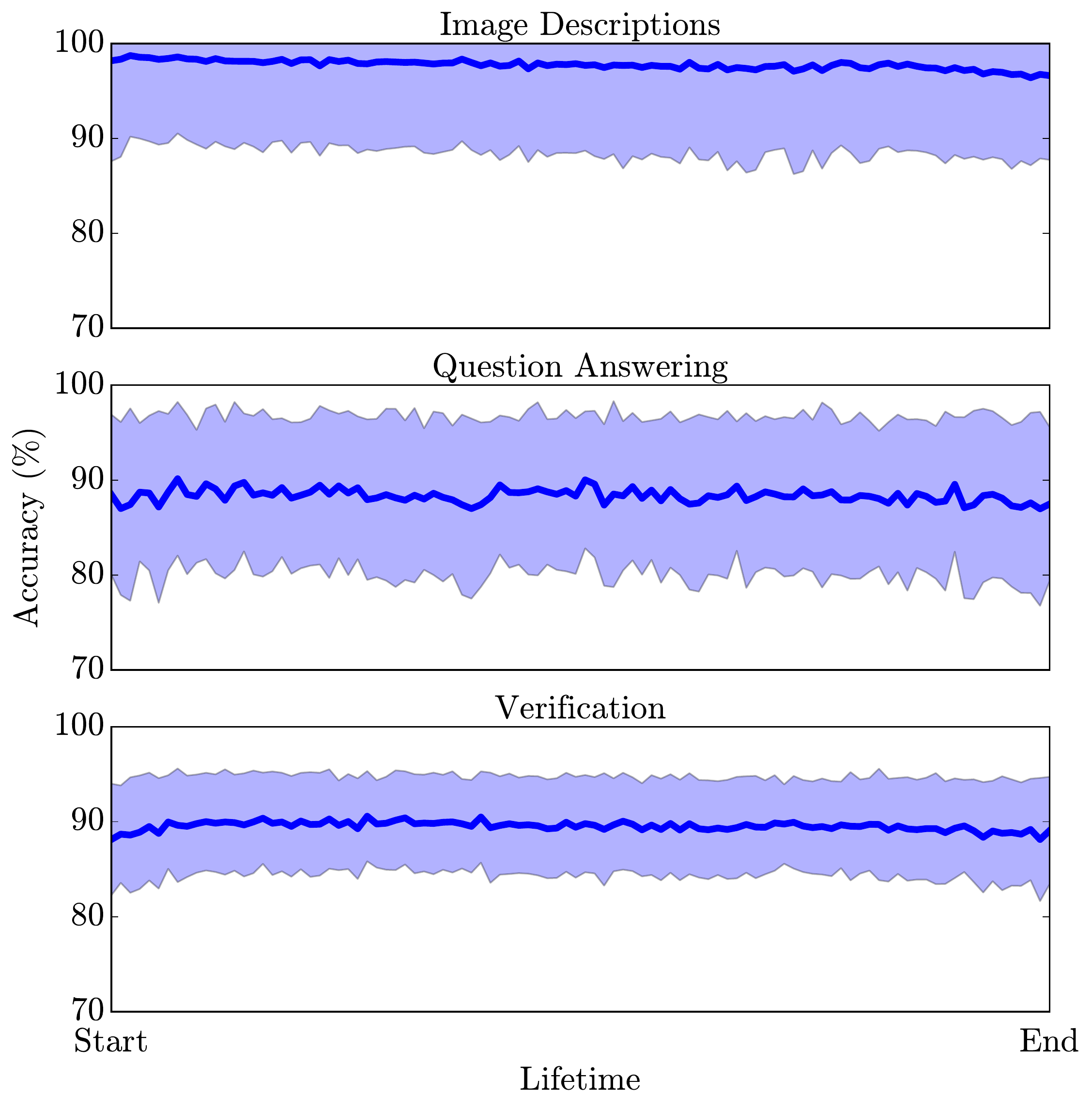}
    \caption{The average accuracy over the lifetime of each worker who completed over $100$ tasks in each of the three datasets. The top row shows accuracy for image descriptions, the middle row shows accuracy for question answering, and the bottom row shows accuracy for the verification dataset. }
    \label{fig:correctness}
\end{figure}
\begin{figure}
    \centering
    \includegraphics[width=\columnwidth]{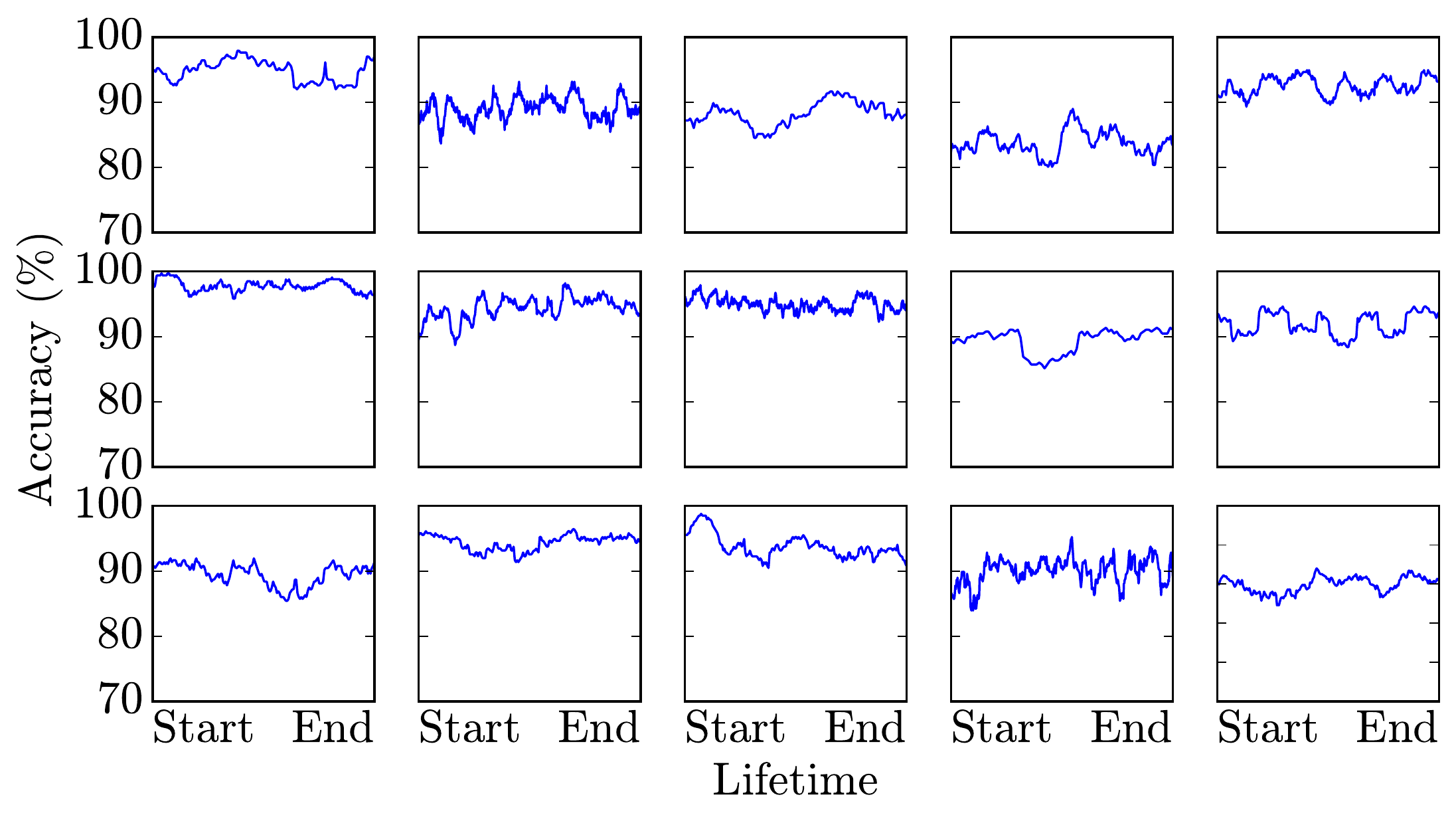}
    \caption{A selection of individual workers' accuracy over time during the question answering task. Each worker remains relatively constant throughout his or her entire lifetime. }
    \label{fig:qa_individuals}
\end{figure}

We find that workers change very little over time (Figures~\ref{fig:correctness} and~\ref{fig:qa_individuals}). When considering those who did at least 100 image description tasks, people on average started at $97.9\pm12.1\%$  accuracy and ended at $96.6\pm9.1\%$, averaging an absolute change of $3.3\pm5.6\%$. Workers who did at least $100$ question answering tasks started with an average of $88.4\pm6.3\%$ and ended at $87.5\pm6.0\%$, resulting in an absolute change of $3.1\pm3.3\%$. For the verification task, workers agreed with the majority on average $88.1\pm3.6\%$ at the start and $89.0\pm4.0\%$ at the end, resulting in an absolute change of $3.1\pm3.4\%$.

\second{Annotation diversity.}
Accuracy captures clearly correct or incorrect outcomes, but how about subtler signals of effort level?
Since each image description or question answering task produces multiple phrases or questions, we examine the linguistic similarity of these phrases and questions over time. As N-grams have often been used in language processing for gauging similarity between documents~\cite{damashek1995gauging}, we construct a metric of syntax diversity for a set of annotations as follows:
\begin{equation}
    \text{diversity} = \frac{\text{number of unique N-grams}}{\text{number of total N-grams}}.
\end{equation}
As the annotation set increasingly contains different words and ordering of words, this diversity metric approaches $1$ because the number of unique N-grams will approach the total possible N-grams. Conversely, if the annotation set contains increasingly similar annotations, many N-grams will be redundant, making this diversity metric approach $0$. To account for workers reusing similar sentence structure in consecutive tasks, we track the number of unique N-grams versus total N-grams in sequential pairs of tasks.

\begin{figure}[t]
    \centering
    \includegraphics[width=\columnwidth]{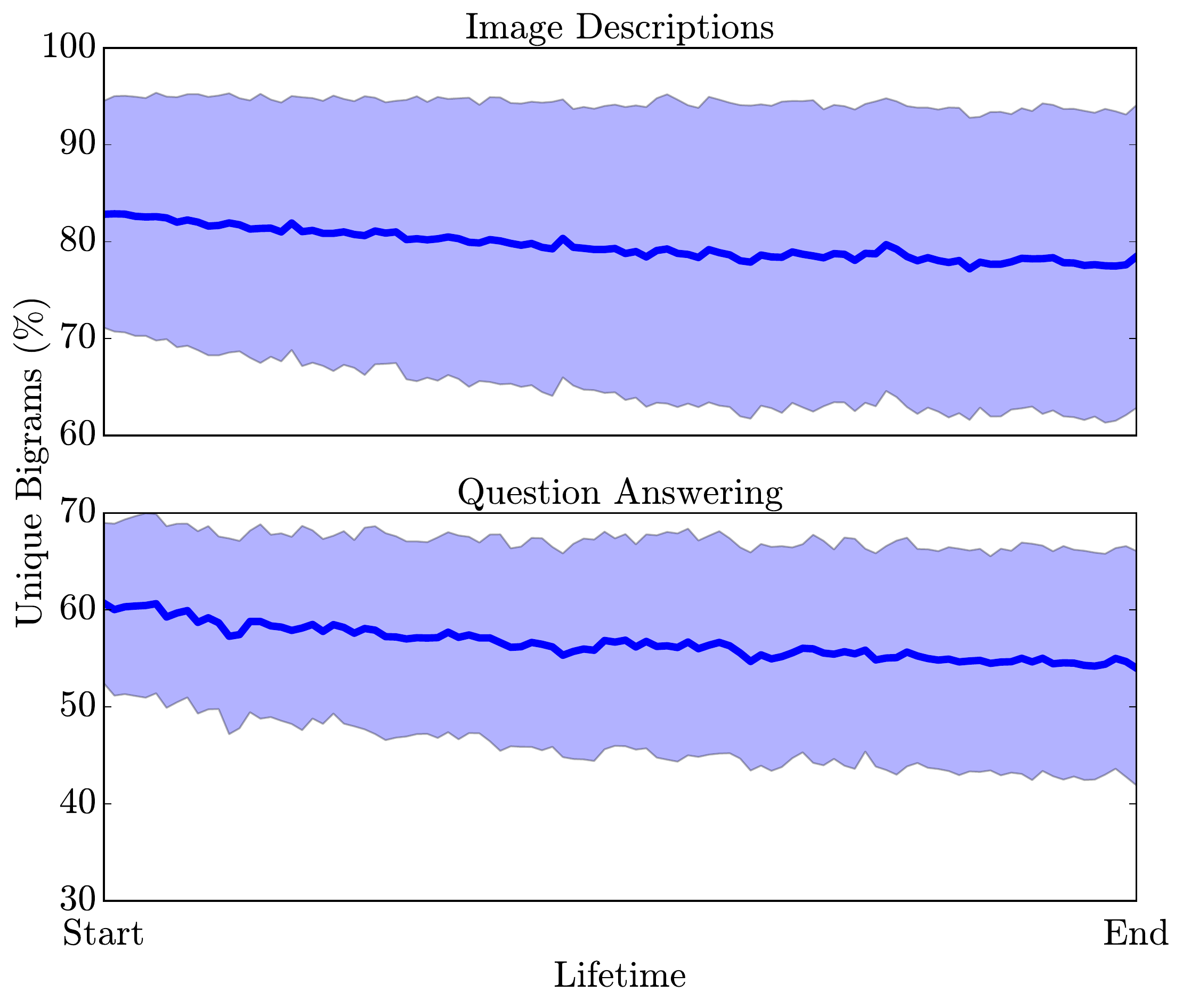}
    \caption{On average, workers who repeatedly completed the image description (top row) or question answering (bottom row) tasks gave descriptions or questions with increasingly similar syntactic structures.}
    \label{fig:bigram}
\end{figure}

Figure~\ref{fig:bigram} illustrates that the percentage of unique bigrams decreases slightly over time. In the image description task, the percent of unique bigrams decreases on average from $82.4\%$ to $78.4\%$ between the start and end of a worker's lifetime. Since there are $4.2$ bigrams on average per phrase, a worker writes approximately $42$ total bigrams per task. Thus, a decrease in $4.0\%$ results in a loss of $1.7$ unique bigrams per task. In the question answering task, the percent of unique bigrams decreases on average from $60.7\%$ to $54.0\%$. As there are on average $3.4$ bigrams per question, this $6.7\%$ decrease would cost a loss of $3.2$ distinct bigrams per task. Ultimately, these results show that over the course of a worker's lifetime, only a small fraction of diversity is lost, as less than a sentence or question's contribution of bigrams is lost.

\begin{figure}[t]
    \centering
    \includegraphics[width=\columnwidth]{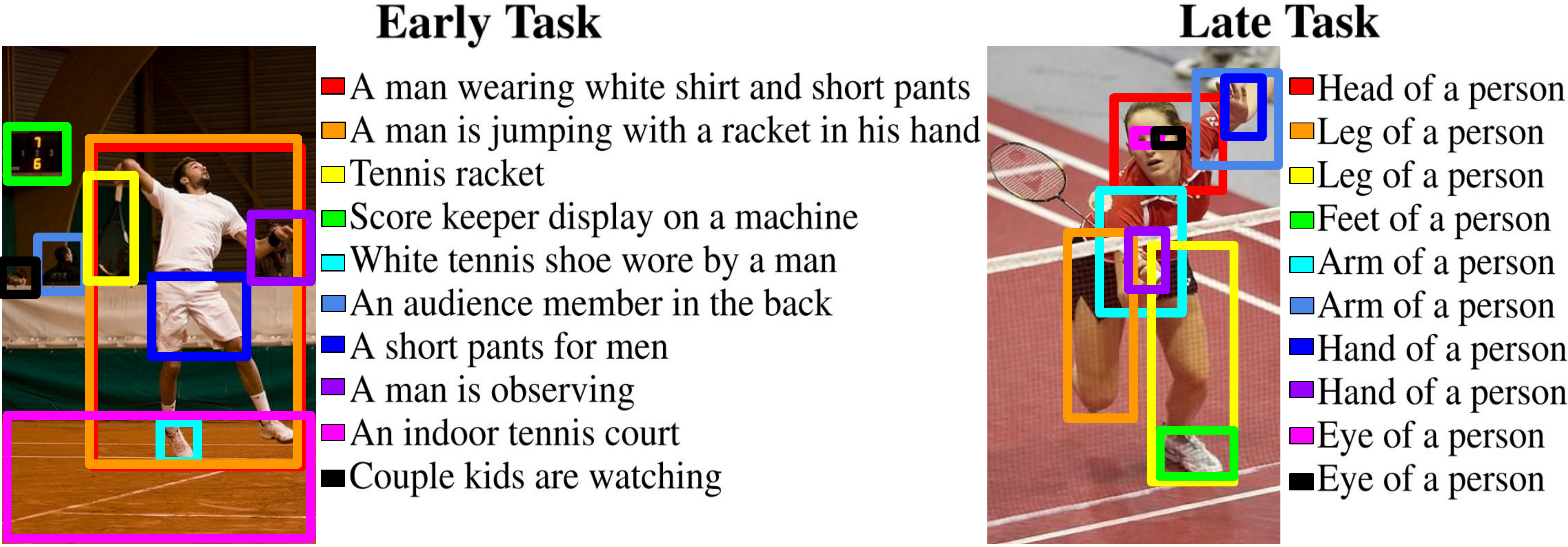}
    \caption{Image descriptions written by a satisficing worker on a task completed near the start of their lifetime (left) and their last completed task (right). Despite the images being visually similar, the phrases submitted in the last task are much less diverse than the ones submitted in the earlier task.}
    \label{fig:both_bbox}
\end{figure}

A majority of workers stay constant during their lifetime. However, a few workers decrease to an extremely low N-gram diversity, despite writing factually correct image descriptions and questions. This behavior describes a ``satisficing'' worker, as they repeatedly write the same types of sentences or questions that generalize to almost every image. Figure~\ref{fig:both_bbox} demonstrates how a satisficing worker's phrase diversity decreases from image-specific descriptions submitted in early-lifetime tasks to generic, repeated sentences submitted in late-lifetime tasks. To determine the percentage of total workers who are satisficing workers, we first compute the average diversity of submissions for each worker. We then set a threshold equal the difference between the maximum and mean of these diversities, labeling workers below the mean by this threshold as satisficers. We find that approximately $7\%$ and $2\%$ of workers satisfice in the image description and question answering datasets respectively. 

\begin{figure}[t]
    \centering
    \includegraphics[width=\columnwidth]{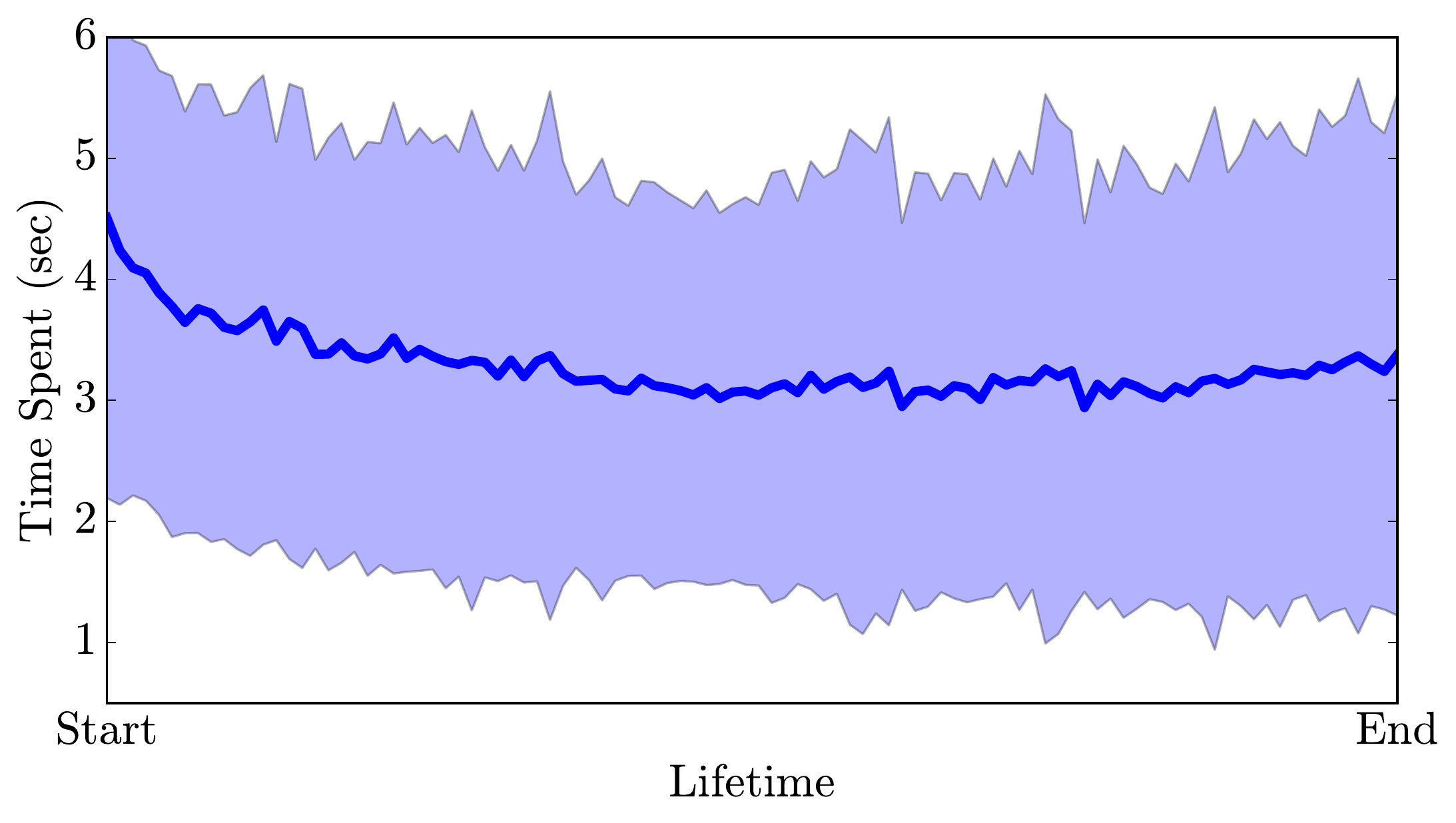}
    \caption{As workers gain familiarity with a task, they become faster. Verification tasks speed up by 25\% from novice to experienced workers.}
    \label{fig:qa_ver_time}
\end{figure}

\second{Annotation speed.}
We recorded the time it takes on average for workers to complete a single verification. We removed $2.4\%$ of the data points deemed as outliers from this computation, as workers will infrequently take longer times during a break or while reading the instructions. We defined outliers for each task of 50 verifications as times that outside 3 standard deviations of the mean time for those 50 verifications. Overall, Figure~\ref{fig:qa_ver_time} demonstrates that workers indeed get faster over time. Initially, workers start off taking $4.5$ seconds per verification task, but end up averaging under $3.4$ seconds per task, resulting in an approximate $25\%$ speedup. Although no time data was recorded for either the image descriptions or question answering tasks, we believe that they would also exhibit similar speedups over time due to practice effects~\cite{newell1981mechanisms} and similarities in the correctness and diversity metrics.
    
\first{Discussion}

\second{No significant fatigue effects are exhibited by in long-term workers.} Workers do not appear to suffer from long-term fatigue effects. With an insignificant average accuracy drop of on average $1.5\%$ for workers across their lifetime, we find that workers demonstrate little change in their submission quality. Instead of suffering from fatigue, workers may be opting out or breaking whenever they feel tired~\cite{dai2015and}. Furthermore, this finding agrees with previous literature that cumulative fatigue is not a major factor in quality drop~\cite{perelli1980fatigue}.

\second{Accuracy is constant within a task type, but varies across different task types.} We attribute the similarity between the average accuracy of the question answering and verification tasks to their sequential relationship in the crowdsourcing pipeline. If the question-answer pairs are ambiguous or speculative, then the majority vote often becomes split, resulting in accuracy loss for both the question answering and verification tasks.  Additionally, we notice the average accuracy for image descriptions is noticeably higher than the average accuracy for either the question answering or verification datasets. We believe this discrepancy stems from the question answering task's instructions that ask workers to write at four distinct types of \emph{W} questions (e.g. ``why'', ``what'', ``when''). Some question types such as ``why'' or ``when'' are often ambiguous for many images (e.g. ``why is the man angry?''). Such questions are often marked as incorrect by other workers in the verification task. Furthermore, we also attribute the disparity between unique bigram percentage for the image description and question answering tasks to the question answering task's instructions that asked workers to begin each question with one of the $7$ question types.

\second{Experience translates to efficiency.} Workers retain constant accuracy, and slightly reduce the complexity of their writing style. Combined, these findings suggest that workers find a general strategy that leads to acceptance and stick with it. Studies of practice effects suggest that a practiced strategy helps to increase worker throughput according to a power law~\cite{newell1981mechanisms}. This power law shape is clearly evident in the average verification speed, confirming that practice plays a crucial role in the worker speedup.

\second{Overall findings.} 
From an analysis of the three datasets, we found that fatigue effects are not significantly visible and that severe satisficing behavior only affects a very small proportion of workers. On average, workers maintain a similar quality of work over time, but also get more efficient as they gain experience with the task. 

\section{Experiment: Why Are Workers Consistent?}
Examining the image descriptions, question answering, and the verification datasets, we find that worker's performance on a given microtask remains consistent --- even if they do the task for multiple months. However, mere observation of this consistency does not give true insight into the reasons for its existence. Thus, we seek to answer the following question: do crowd workers satisfice according to the minimum quality necessary to get paid, or are they consistent regardless of this minimum quality? 

To answer this question, we perform an experiment where we vary the quality threshold of work and the threshold's visibility. If workers are stable, we would expect them to either submit work that is above or below the threshold, irrespective of what the threshold is. However, if workers satisfice according to the minimum quality expected, they would adjust the quality of their work based on set threshold~\cite{simon1972theories,krosnick1991response}. 

If workers indeed satisfice, then the knowledge of this threshold and their own performance should make it easier to perfect satisficing strategies. Therefore, to adequately study the effects of satisficing, we vary the visibility of the threshold to workers as well. In one condition, we display workers' current quality scores and the minimum quality score to be accepted, while the other condition only displays whether submitted work was accepted or rejected. To sum up, we vary the threshold and the transparency of this threshold to determine how crowd workers react to the same task, but with different acceptability criteria.

\begin{figure}[t]
    \centering
    \includegraphics[width=\columnwidth]{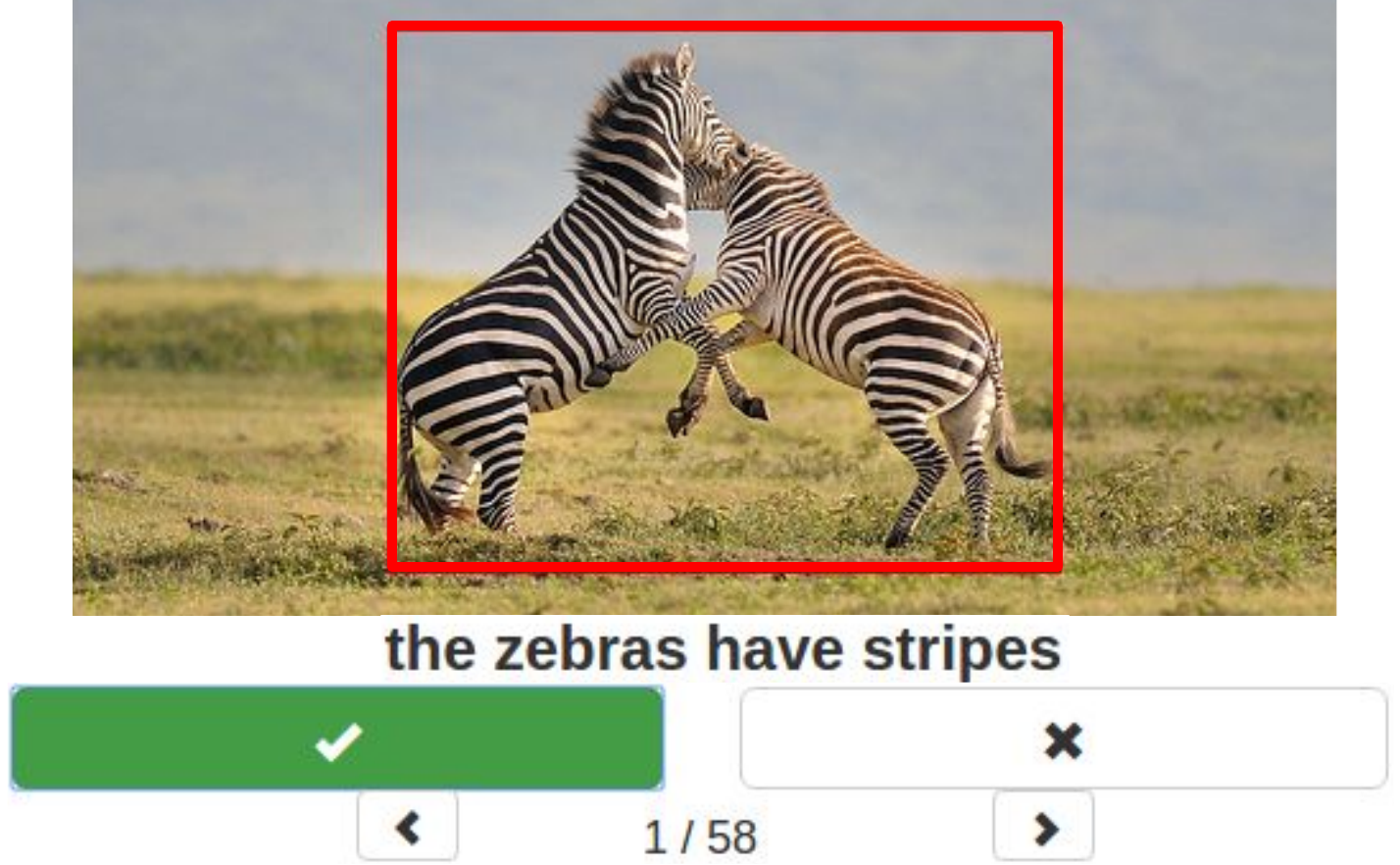}
    \caption{An example binary verification task where workers are asked to determine if the phrase ``the zebras have stripes'' is a factually correct description of the image region surrounded within the red box. There were $58$ verification questions in each task.}
    \label{fig:verification_task}
\end{figure}

\first{Task}
To study why workers are consistent, we designed a task where workers are presented with a series of randomly ordered $58$ binary verification questions. Each verification requires them to determine if an image description and its associated image part are correct. For example, in Figure~\ref{fig:verification_task}, workers must decide if ``the zebras have stripes'' is a good description of a particular part of the image. They are asked to base their response based solely on the content of the image and the semantics of the sentence. To keep the task simple, we asked workers to ignore whether the box was perfectly surrounding the image area being described. The tasks were priced such that workers could earn $\$6$ per hour and were available to workers with a $95\%$ approval rating and who lived in the United States. Each task took approximately $4$ minutes to complete and were given $2.5$ hours to complete the task to ensure workers were not pressured for time. 

\begin{figure}[t!]
    \centering
    \includegraphics[width=\columnwidth]{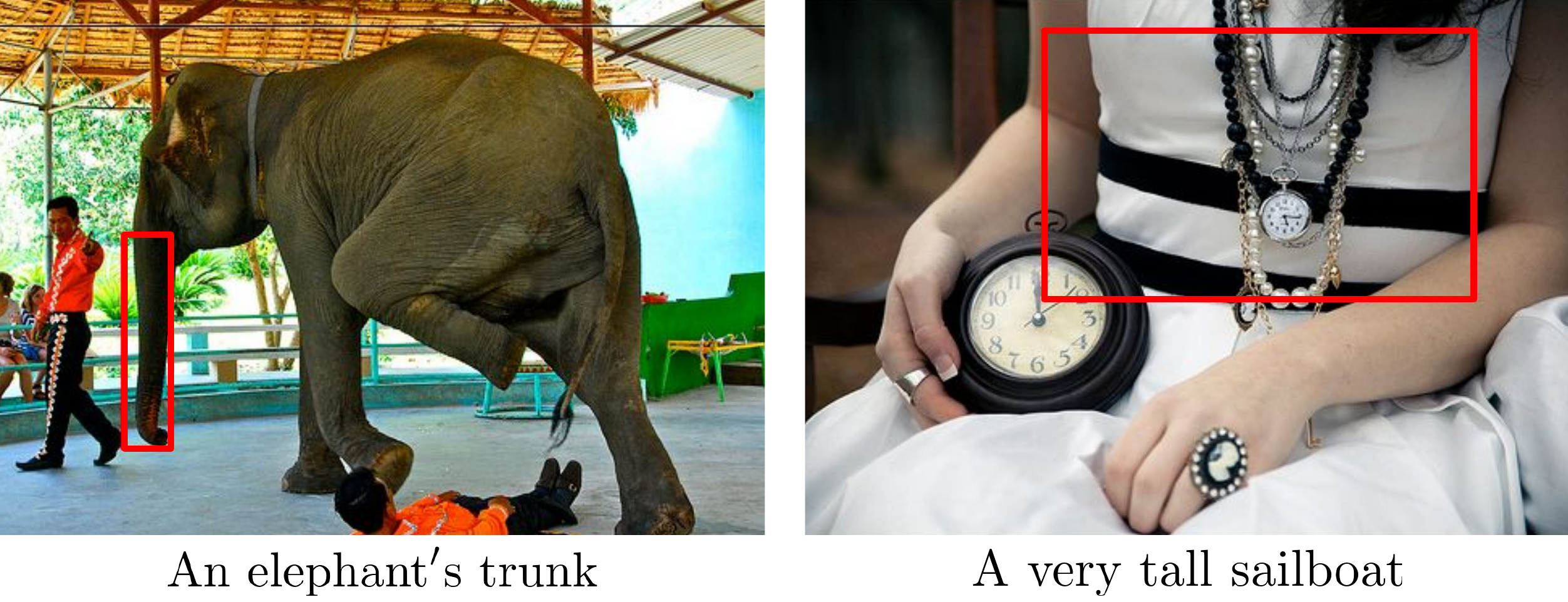}
    \caption{Examples of attention checks placed in our binary verification tasks. Each attention check was designed such that they were easily identified as correct or incorrect. ``An elephant's trunk'' (left) is a positive attention check while ``A very tall sailboat'' (right) is an incorrect attention check. We rated worker's quality by measuring how well they performed on these attention checks.}
    \label{fig:gold}
\end{figure}

We placed $3$ attention checks in each task. Attention checks are gold-standard verification questions whose answers were already known. Attention checks were randomly placed within the series of $58$ verifications to gauge how well a worker performed on the given task. To avoid workers from incorrectly marking an attention check due to subjective interpretation of the description, we manually marked these attention checks correct or incorrect. Examples of attention checks are shown in Figure~\ref{fig:gold}. Incorrect attention checks were completely mismatched from their image; for example ``A very tall sailboat'' was used as a incorrect attention check matched to an image of a lady wearing a white dress. We created a total of $11,290$ unique attention checks to prevent workers from simply memorizing the attention checks.

Even though these attention checks were designed to be obviously correct or incorrect, we ensured that we do not reject a worker's submission based off a single, careless mistake or an unexpected ambiguous attention check. After completing a task, each worker's submission is immediately accepted or rejected based on a rating, which is calculated as the percentage of the last $30$ attention checks correctly labeled. If a worker's rating falls below the threshold of acceptable quality, their task is rejected. However, to ensure fair payment, even if a worker's rating is below the threshold, their task is accepted if they get all the attention checks in the current task correct. This enables workers who are below the threshold to perform carefully and improve their rating as they continue to do more tasks.

\first{Experiment Setup}
Our goal is to vary the acceptance threshold to see how it impacts worker quality over time. We performed a between-subjects $2\times2$ study where we varied \textit{threshold} and \textit{transparency}. We ran an initial study with a different set of $100$ workers to estimate how people performed on this verification task. We found that workers get a mean accuracy of $94\pm10\%$ with a median accuracy of $95.5\%$. We chose the thresholds such that the high threshold condition asked workers to perform above the median and the low threshold was below $2\times$ the standard deviation, allowing workers plenty of room to make mistakes. The  high threshold factor level was set at $96\%$ while the low threshold factor level was set at $70\%$. Workers in the high threshold level could only incorrectly label at most $1$ out of $30$ of the previous attention checks to avoid rejection, while workers in the low threshold level could error on $8$ out of the past $30$ attention checks.

We used two levels of transparency: high and low. In the high factor level, workers were able to see their current rating at the beginning of every task and were also alerted of how their rating changed after submitting each task. Meanwhile, in the low factor level, workers did not see their rating, nor did they know what their assigned threshold was.

We recruited workers from AMT for the study and  randomized them between conditions. We measured workers' accuracy and total number of completed tasks under these four conditions.

{
\first{Data Collected}
By the end of the study, $1,134$ workers completed $\num[group-separator={,}]{11666}$ tasks. In total, $\num[group-separator={,}]{676628}$ binary verification questions were answered, of which $\num[group-separator={,}]{34998}$ were attention checks.  Table~\ref{tab:consistent_data} shows the breakdown of the number of workers who completed at least $1$ task. Not all workers who accepted tasks completed them. In the high threshold condition, $106$ and $116$ workers did not complete any tasks in the high and low transparency conditions respectively. Similarly, $137$ and $138$ workers did not complete tasks in the low threshold. This resulted in $29$ and $28$ more people in the low threshold that completed tasks. Workers completed on average a total of $576$ verifications each. 

\first{Results}
On average, the accuracy of the work submitted by workers in all  four conditions remained consistent (Figure~\ref{fig:consistency_plots}). In the low threshold factor level, workers averaged a rating of $93.6\pm4.7\%$ and $93.3\pm5.8$ in the high and low transparency factor levels. Meanwhile, when the threshold was high, workers in the low transparency factor level averaged $94.2\pm4.4\%$ while the workers in the high transparency factor level averaged $95.2\pm4.0\%$.  Overall, the high transparency factor level had a smaller standard deviation throughout the course of workers' lifetimes. We conducted a two-way ANOVA using the two factors as independent variables on all workers who performed more than $5$ tasks. The ANOVA found that there was no significant effect of threshold (F($1$, $665$)=$0.55$, p=$0.45$) or transparency (F($1$, $665$)=$2.29$, p=$0.13$), and no interaction effect (F($1$, $665$)=$0.24$, p=$0.62$). Thus, worker accuracy was unaffected by the accuracy requirement of the task.

Unlike accuracy, worker retention was influenced by our manipulation. By the $50^{th}$ task, less than $10\%$ of the initial worker population continued to complete tasks.  This result is consistent with our observations with the Visual Genome datasets and from previous literature that explains that a small percentage of workers complete most of the crowdsourced work~\cite{little8020}. We also observe that workers in the high threshold and high transparency condition have a sharper dropout rate in the beginning. To measure the effects of the four conditions on dropout, we analyzed the logarithm of the number of tasks completed per condition using an ANOVA. (Log-transforming the data ensured that it was normally distributed and thus amenable to ANOVA analysis.) The ANOVA found that there was a significant effect of transparency (F($1$, $665$)=$279.87$, p<$0.001$) and threshold (F($1$, $665$)=$88.61$, p<$0.001$), and also a significant interaction effect (F($1$, $665$)=$76.23$, p<$0.001$). A post hoc Tukey test~\cite{tukey1949comparing} showed that the (1) high transparency and high threshold condition had significantly less retention than the (2) low transparency and high threshold condition ($p < .05$).

\begin{table}[t]

\centering
    \begin{tabular}{l | c c | c c }
    Threshold &
      \multicolumn{2}{c|}{High: 96} &
      \multicolumn{2}{c}{Low: 70} \\
    Transparency & High & Low & High & Low \\
    \hline
    \rule{0pt}{2ex}\# workers with 0 tasks & 106 & 116 & 137 & 138 \\
    \# workers with >1 tasks & 267 & 267 & 300 & 300 \\
    \# tasks & 2702 & 2630 & 3209 & 3125 \\
    \# verifications & 5076 & 7890 & 9627 & 9375 \\
  \end{tabular}
\caption{Data collected from the verification experiment. A total of $1,134$ workers were divided up into four conditions, with a high or low threshold and transparency.}
\label{tab:consistent_data}
\end{table}

\begin{figure*}[t]
    \centering
    \begin{subfigure}[b]{0.48\textwidth}
        \includegraphics[width=\textwidth]{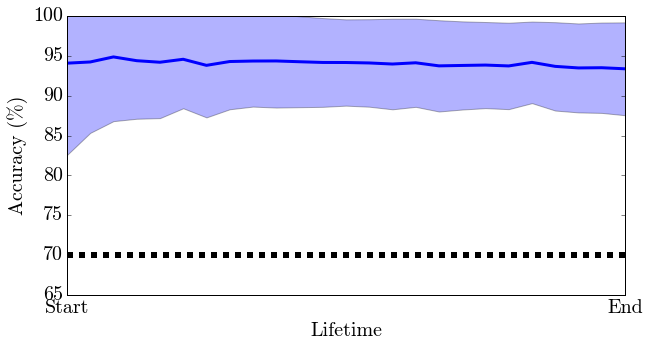}
        \caption{Threshold: low ($70$) and transparency: low, accuracy: $93.3\pm5.8\%$}
        \label{fig:ul}
    \end{subfigure}
    ~ 
    \begin{subfigure}[b]{0.48\textwidth}
        \includegraphics[width=\textwidth]{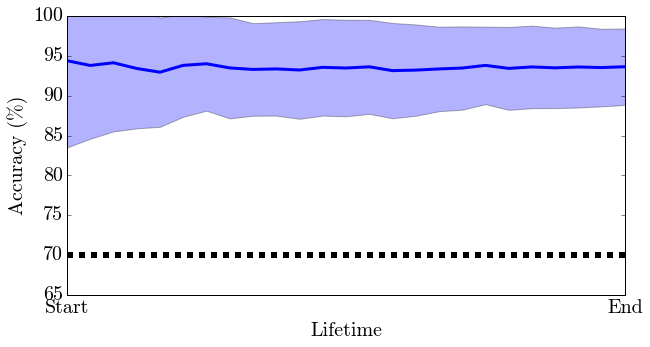}
        \caption{Threshold: low ($70$) and transparency: high, accuracy: $93.6\pm4.7\%$}
        \label{fig:kl}
    \end{subfigure}
    ~ 
    \begin{subfigure}[b]{0.48\textwidth}
        \includegraphics[width=\textwidth]{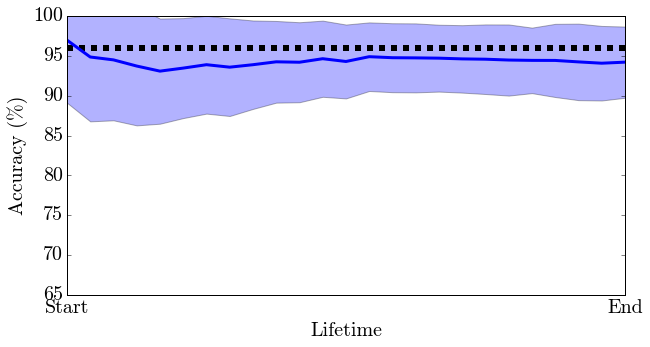}
        \caption{Threshold: high ($96$) and transparency: low, accuracy: $94.2\pm4.4\%$}
        \label{fig:uh}
    \end{subfigure}
    \begin{subfigure}[b]{0.48\textwidth}
        \includegraphics[width=\textwidth]{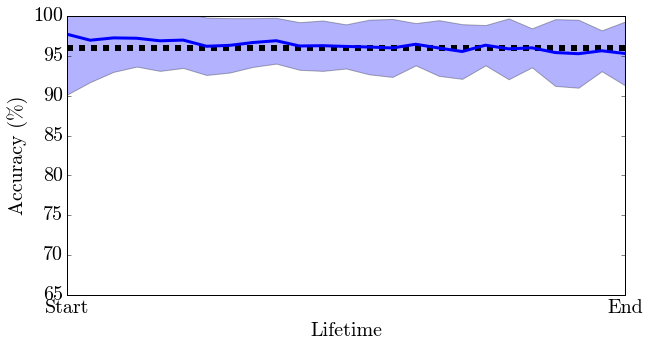}
        \caption{Threshold: high ($96$) and transparency: high, accuracy: $95.2\pm4.0\%$}
        \label{fig:kh}
    \end{subfigure}
    \caption{Worker accuracy was unaffected by the threshold level and by the visibility of the threshold. The dotted black line indicates the threshold that the workers were supposed to adhere to.}
    \label{fig:consistency_plots}
\end{figure*}

\first{Discussion}
\second{Workers are consistent in their quality level.} With this experiment, we are now ready to answer whether workers are consistent or satisficing to an acceptance threshold. Given that workers' quality was consistent throughout all the four conditions, evidence suggests that workers were consistent, regardless of the threshold at which requesters accept their work. In the low threshold and high transparency condition, workers are aware that their work will be accepted if their rating is above $70$\%, and still perform with an average rating of $94$\%. Workers are risk-averse, and seek to avoid harms to their acceptance rate~\cite{mcinnis2016taking}. Once they find a strategy that allows their work to be accepted, they stick to that strategy throughout their lifetime~\cite{mitra2015comparing}. This result is consistent with the earlier observational data analysis.

\second{Workers minimize risk by opting out of  tasks above their natural accuracy level.} If workers do not adjust their quality level in response to task difficulty, the only other possibility is that workers  self-select out of tasks they cannot complete effectively. Our data supports this hypothesis: workers in the high transparency and high threshold  condition did statistically fewer tasks on average. The workers self-selected  out of the task when they had a higher chance of rejection. Out of $267$ workers in the high transparency and high threshold condition, $200$ workers workers stopped working once their rating dropped below the $96$\% threshold. Meanwhile, in the high transparency and low threshold condition, out of the $300$ workers who completed our tasks, almost all of them continued working even if their rating dropped below the $70$\% threshold, often bringing their rating back up to above $96$\%.

This study illustrates that workers are consistent over very long periods of hundreds of tasks. They quickly develop a strategy to complete the task within the first few tasks and stick with it throughout their lifetime. If their work is  approved, they continue to complete the task using the same strategy. If their strategy begins to fail, instead of adapting, they self-select themselves out of the task. 

\section{Predicting From Small Glimpses}
The longitudinal analysis in the first section and the experimental analysis in the second section found that crowd worker quality remains consistent regardless of how many tasks the worker completes and regardless of the required acceptance criteria. Bolstered by this result, this section demonstrates the efficacy of predicting a worker's future quality by observing a small glimpse of their initial work. The ability to predict a worker's quality on future tasks can help requesters identify good workers and improve the quality of data collected.

\begin{figure}[t]
    \centering
    \includegraphics[width=\columnwidth]{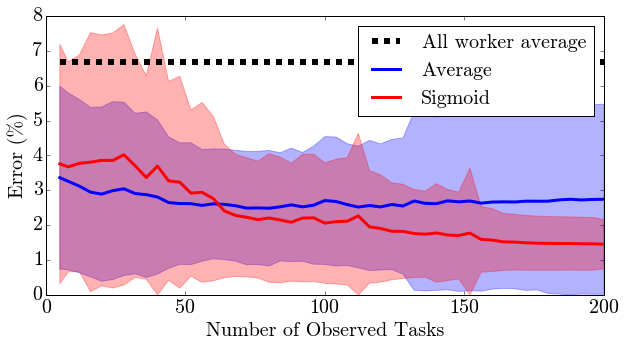}
    \caption{It is possible to model a worker's future quality by observing only a small glimpse of their initial work. Our \textit{all workers' average} baseline assumes that all workers perform similarly and manages an error in individual worker quality prediction of $6.9\%$. Meanwhile, by just observing the first 5 tasks, our \textit{average} and \textit{sigmoid} models achieve $3.4\%$ and $3.7\%$ prediction error respectively. As we observe more hits, the \textit{sigmoid} model is able to represent workers better than the \textit{average} model.}
    \label{fig:model}
\end{figure}

\first{Experimental Setup} 
To create a prediction model, we use the question answering dataset. Our aim is to predict a worker's quality on the task towards the end of their lifetime. Since workers' individual quality on every single task can be noisy, we estimate a worker's future quality as the average of their accuracy on the last $10\%$ of their tasks in their lifetime.

We allow our model to use between the first $5$ and the first $200$ tasks completed by a worker to estimate their future quality. Therefore, we only test our model on workers who have completed at least $200$ tasks. As a baseline, we calculate \textit{the average of all workers'} performances on their last $n$ tasks. We use this value as our guess for each individual worker's future quality. This model assumes a worker does as well as the average worker does on their final tasks.

Besides the baseline, we use two separate models to estimate a worker's future quality: \textit{average} and \textit{sigmoid} models. The average model is a simple model that uses the average of the worker's $n$ tasks as the estimate for all future quality predictions. For example, if a worker averages $90$\% accuracy on their first five tasks, the average model would predict that the worker will continue to perform at a $90$\% accuracy. However, if the worker's quality on their last $10$\% of tasks is $85$\%, then the prediction error would be $5$\%. The \textit{sigmoid} model attempts to represent a worker's quality as a sigmoid curve with $4$ parameters to adjust for the offset of the curve. We use a sigmoid model because we find that many workers display a very brief learning curve over the first few tasks and remain consistent thereafter. The initial adjustment and future consistency closely resembles a sigmoid curve.

\first{Results} 
The \textit{average of all workers'} accuracy is $87.8\%$. Using this value as a baseline model for quality yields an error of $6.9\%$. We plot the error of the baseline as a dotted line in Figure~\ref{fig:model}. The \textit{average} model performs better: even for only a glimpse of $n=5$ tasks, its error is $3.4\%$. After seeing a worker's first $n=200$ tasks, the model gets slightly better and has a prediction error of $3.0\%$. The \textit{sigmoid} model outperforms the baseline but underperforms the average model and achieves an error of $3.7\%$ for $n=5$. As the model incorporates more tasks, it becomes the most accurate, managing an error rate of $1.4\%$ after seeing $n=200$ tasks. Furthermore, the model's standard deviation of the error also decreases from $3.4\%$ to $0.7\%$ as $n$ increases.

\first{Discussion}
\textit{Even a glimpse of five tasks can predict a worker's future quality.} Since workers are consistent over time, both the \textit{average} and the \textit{sigmoid} models are able to model workers' quality with very little error. When workers initially start doing work, a simple \textit{average} model is a good choice for a model to estimate how well the worker might perform in the future. However, as the worker completes more and more tasks, the \textit{sigmoid} model is able to capture the initial adjustment a worker makes when starting a task. By utilizing such models, requesters can estimate which workers are most likely to produce good work and can easily qualify good workers for long-term work.

\section{Implications for Crowdsourcing}
\textit{Encouraging diversity.} The consistent accuracy and constant diversity of worker output over time makes sense from a practical perspective: workers are often acclimating to a certain style of completing work~\cite{mcinnis2016taking} and often adopt a particular strategy to get paid. However, this formulaic approach might run counter to a requester's desire to have richly diverse responses. Checks to increase diversity, such as enforcing a high threshold for diversity, should be employed without fear of worker quality as we have observed that quality does not significantly change with varying acceptance thresholds. Therefore, designing tasks that promote diversity without effecting the annotation quality is a ripe area for future research.

\textit{Worker retention.} Additional experience affects completion speeds but does not translate to higher quality data. Much work has been done to retain workers \cite{dai2015and,difallah2014scaling,law2016curiosity}, but, as shown, retention does not equate to increases in worker quality --- just more work completed. Further work should be conducted to not only retain a worker pool, but also examine methods of identifying good workers~\cite{karger2011budget} and more direct interventions for training poorly performing workers~\cite{dow2012shepherding,kittur2008crowdsourcing}. 

Additionally, other studies have shown that the motivation of workers is the predominant factor in the development of fatigue, rather than the total time worked~\cite{beckers2004working}. Although crowdsourcing can be intrinsically motivated~\cite{von2006games}, the microtask paradigm found in the majority of crowdsourcing tasks favors a structure that is efficient~\cite{krishna2016embracing,heer2010crowdsourcing} for workers rather than being interesting for them~\cite{chandler2013breaking}. Future tasks should consider building continuity in their workflow design for both individual worker efficiency~\cite{lasecki2014using} and overall throughput and retention~\cite{dai2015and}.

\textit{Person-centric versus process-centric crowdsourcing.} Attaining high quality judgments from crowd workers is often seen as a challenge~\cite{rashtchian2010collecting,shaw2011designing,sorokin2008utility}. This challenge has catalyzed studies suggesting quality control measures that address the problem of noisy or low quality work~\cite{downs2010your,kittur2008crowdsourcing,mason2010financial}. Many of these investigations study various quality-control measures as standalone intervention strategies. While we explored process-centric measures like varying the acceptance or transparency threshold, previous work has experimented with varying financial incentives~\cite{mitra2015comparing}. All the results support the conclusion that process-centric strategies do not have significant difference in the quality of work submitted. While we agree that such process focused strategies are important to explore, our data reinforces that person-centric strategies (like utilizing worker approval ratings or worker quality on initial tasks) may be more effective~\cite{mitra2015comparing,rzeszotarski2012crowdscape} because they identify a worker's (consistent) quality early on. 

\second{Limitations.} Our analysis solely focuses on data labeling microtasks, and we have not yet studied whether our findings translate over to more complex tasks, such as designing an advertisement or editing an essay~\cite{kittur2011crowdforge,bernstein2010soylent}. Furthermore, we focus on weeks-to-months crowd worker behavior based on datasets collected over a few months, but there exist some crowdsourcing tasks~\cite{brelig2013system} that have persisted far longer than our study. Thus, we leave the analysis of crowd worker behavior spanning multiple years to future work.


\section{Conclusion}
Microtask crowdsourcing is rapidly being adopted to generate large datasets with millions of labels. Under the Pareto principle, a small minority of workers complete a great majority of the work. In this paper, we studied how the quality of workers' submissions change over extended periods of time as they complete thousands of tasks. Contrary to previous literature on fatigue and satisficing, we found that workers are extremely consistent throughout their lifetime of submitting work. They adopt a particular strategy for completing tasks and continue to use that strategy without change. To understand how workers settle upon their strategy, we conducted an experiment where we vary the required quality for large crowdsourcing  tasks. We found that workers do not satisfice and consistently perform at their usual quality level. If their natural quality level is below the acceptance threshold, workers tend to opt out from completing further tasks. Due to this consistency, we demonstrated that brief glimpses of just the first five tasks can predict a worker's  long-term quality. We argue that such consistent worker behavior must be utilized to develop new crowdsourcing strategies that find good workers and collect unvarying high quality annotations.

\textit{Acknowledgements.} Our work is partially funded by an ONR MURI grant.

\bibliographystyle{SIGCHI-Reference-Format}
\bibliography{sample}

\end{document}